\documentclass[conference]{IEEEtran}
\IEEEoverridecommandlockouts
\usepackage{cite}
\usepackage{amsmath,amssymb,amsfonts}
\usepackage{algorithmic}
\usepackage{tabularx}
\usepackage{multirow}
\usepackage{graphicx}
\usepackage{subcaption}
\usepackage{textcomp}
\usepackage{flushend}
\usepackage[table,xcdraw]{xcolor}
\def\BibTeX{{\rm B\kern-.05em{\sc i\kern-.025em b}\kern-.08em
    T\kern-.1667em\lower.7ex\hbox{E}\kern-.125emX}}
\begin{document}

\title{Comparing Adversarial and Supervised Learning for  Organs at Risk Segmentation in CT images\\
\thanks{This paper is supported by the PNRR-PE-AI FAIR project funded by the NextGeneration EU program.}}

\author{
\IEEEauthorblockN{Leonardo Crespi\IEEEauthorrefmark{4}\IEEEauthorrefmark{2}, Mattia Portanti\IEEEauthorrefmark{4}, Daniele Loiacono\IEEEauthorrefmark{4}}
\smallskip
\IEEEauthorblockA{\IEEEauthorrefmark{4}Dipartimento di Elettronica, Informazione e Bioingegneria, Politecnico di Milano, Milan, Italy}
\IEEEauthorblockA{\IEEEauthorrefmark{2}Centre for Health Data Science, Human Technopole, Milan, Italy}}

\maketitle

\begin{abstract}
Organ at Risk (OAR) segmentation from CT scans is a key component of the radiotherapy treatment workflow. 
In recent years, deep learning techniques have shown remarkable potential in automating this process. 
In this paper, we investigate the performance of Generative Adversarial Networks (GANs) compared to supervised learning approaches for segmenting OARs from CT images. 
We propose three GAN-based models with identical generator architectures but different discriminator networks. These models are compared with 
well-established CNN models, such as SE-ResUnet and DeepLabV3, using the StructSeg dataset, which consists of 50 annotated CT scans containing contours of six OARs. 
Our work aims to provide insight into the advantages and disadvantages of adversarial training in the context of OAR segmentation. 
The results are very promising and show that the proposed GAN-based approaches are similar or superior to their CNN-based counterparts, particularly when segmenting more challenging target organs.
\end{abstract}


\section{Introduction}
\label{sec:intro}
In the medical field, segmentation of regions of interest (ROIs), such as organs, tissues, tumors, or other structures, from various imaging modalities like Computed Tomography (CT), Magnetic Resonance Imaging (MRI), and X-ray scans, plays a crucial role in numerous diagnostic and treatment workflows, particularly in radiotherapy. 
Manual segmentation can be challenging, prone to error, and extremely time-consuming, especially when handling large or multiple targets. 
Consequently, automating this process has garnered significant interest within the scientific community and might lead to substantial improvements in clinical workflows.

A notable application in radiotherapy involves the segmentation of Organs at Risk (OAR/OARs), which necessitates comprehensive and precise delineation of organs and tissues, alongside the identification of target regions. 
This process is critical for treatments like Total Marrow and Lymphoid Irradiation~\cite{wong2020total} and can take more than ten hours to complete. OAR segmentation involves acquiring a patient's CT scan, which is then analyzed by a clinician to identify not only the target areas for irradiation but also the OARs to be spared. This is possible thanks to modern radiation therapy equipment that enables highly precise and customized treatment delivery, tailored to each patient's unique anatomy and needs.

In the past decade, the rapid rise of Deep Learning (DL) has revolutionized computer vision, with Convolutional Neural Networks (CNNs) becoming widely applied in image processing tasks~\cite{gu2018recent}. State-of-the-art models now utilize CNNs, vision transformers~\cite{liu2021swin}, and other DL models for image semantic segmentation. The literature is abundant with diverse architectures, featuring various modifications, inputs, parameters, depth, and training paradigms. Although several models and algorithms have demonstrated remarkable performance~\cite{isensee2018nnu,wasserthal2022totalsegmentator}, the vast array of approaches makes it difficult to determine the best practices for creating a successful model~\cite{crespi20223d}. 
In the medical imaging field, CNN models for segmentation typically employ U-Net style architectures, use 2D input, and are trained in a supervised manner~\cite{wasserthal2022totalsegmentator}.
This is partly due to the increasing availability of annotated public datasets and collaborations with medical centers that provide them.

While supervised training is currently the most popular choice for training CNNs, Generative Adversarial Networks (GANs)~\cite{goodfellow2020generative} have emerged as a highly promising method for generating high-quality images and performing tasks such as image-to-image translation~\cite{isola2017image}, image super-resolution~\cite{ledig2017photo}, and image restoration. Although adversarial learning has been primarily applied to image generation tasks, recent studies have shown its potential for semantic segmentation tasks~\cite{luc2016semantic,li2021semantic}, even in the medical domain~\cite{giacomello2020brain}. 
However, it remains unclear whether using an adversarial learning paradigm could lead to more reliable and better-performing models. 
In this work, we aim to investigate this aspect by comparing two solutions for OAR segmentation, trained using both supervised and adversarial approaches. 

\subsection{Objectives}
\label{sub:pur}
This work aims to explore the merits and drawbacks of using adversarial training, particularly with GANs, in contrast to a supervised learning approach, specifically in the context of segmenting Organs at Risk (OARs) from CT images.

To achieve this goal, we propose three GAN-based models for OARs segmentation in CT scans, drawing inspiration from the work of Tan et al.~\cite{TAN2021101817}. 
These models share the same generator architecture but employ different discriminator networks.
 We compare these models with established and well-tested CNN models, such as SE-ResUnet [7] and DeepLabV3 [8], using a dataset comprising 50 CT scans from various patients, totaling 3861 images. 
 Furthermore, we explore an approach for multi-class segmentation. 
 The comparison focuses on six OARs and employs two evaluation metrics: Dice Score Coefficient (DSC) and Hausdorff Distance (HD), which are among the most commonly used measures for assessing model performance in similar tasks.    

\subsection{Related Works}
\label{sub:rel}
 The popularity of Deep Learning (DL) has led to numerous attempts to improve the performance of state-of-the-art models, often with success. However, it is more challenging to find studies that isolate the effects of specific aspects of training, particularly in the context of medical images where data availability is limited compared to general-purpose datasets. 

In recent years, GANs have been successfully employed in various medical segmentation tasks, demonstrating their potential in this field.
In~\cite{khosravan2019pan}, the authors focus on pancreas segmentation, achieving a Dice Score of 85.53\% by utilizing 2D projections of 3D data. This approach incorporates more contextual information without directly feeding 3D data to the network, a problem also highlighted in~\cite{crespi20223d}. Xue et al.~\cite{xue2018segan} introduce SeGAN for brain tumor segmentation, outperforming the top model in the BRATS tumor segmentation challenge~\cite{baid2021rsna}. Their approach involves a min-max game between a fully convolutional segmentator and a critic, which learns both local and global features by minimizing a multi-scale loss. SegAN-CAT~\cite{giacomello2020brain} extends SegAN by using multiple MRI modalities to enhance performance and applying transfer learning across these modalities. Zhang et al.~\cite{zhang2017deep} address the issue of unannotated data by training the discriminator of their network to distinguish between images generated from examples with or without annotations. This approach results in a robust system capable of segmenting images in both scenarios. In another study, Jin et al.~\cite{jin2018ct} employ GANs for data augmentation to improve segmentation model performance when data is scarce, a significant challenge in some medical specialties. Similarly, Zhao et al.~\cite{zhao2018craniomaxillofacial} use a GAN-based method to map head MRIs to CT images, enhancing the segmentation of craniomaxillofacial bony structures.

Despite these advancements, there appears to be a lack of studies that fairly compare the performance of adversarial or semi-supervised training with supervised approaches.

\section{Methodology} 
\label{sec:met}
In this paper, we aim at comparing GAN models, trained using an adversarial paradigm, with supervised CNN models from the literature, focusing on the semantic segmentation of medical images. Specifically, we target the segmentation of Organs at Risk (OARs) using the StrutSeg2019 dataset~\cite{structseg}, which is one of the most widely used publicly available annotated medical image datasets. The dataset contains 50 annotated thoracic CT scans from lung cancer patients.

\begin{figure}
    \centering
    \includegraphics[width=\columnwidth]{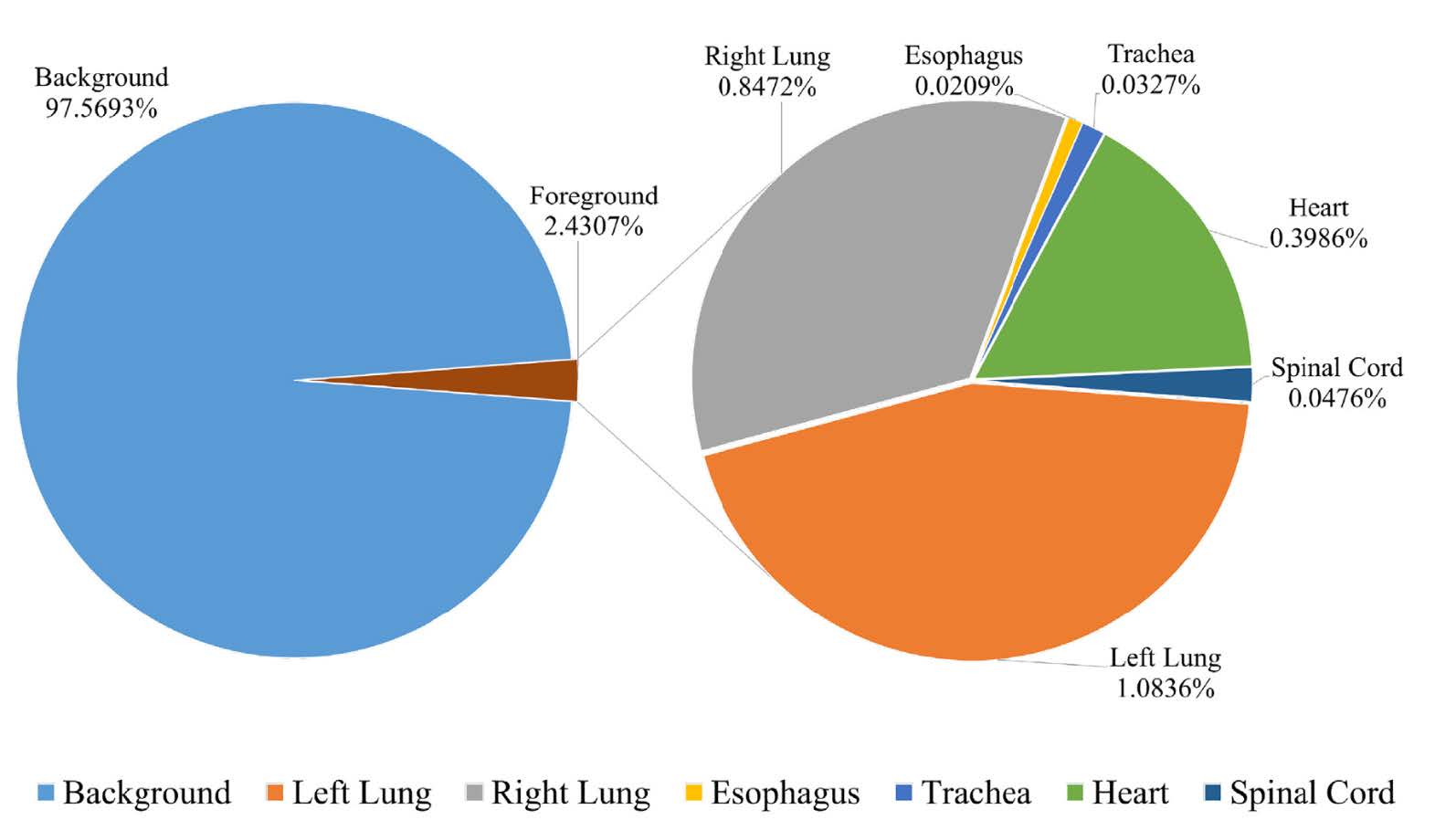}
    \caption{Pixel distribution in the data with respect to the annotated OARs \cite{cao2021cascaded}}
    \label{fig:pie}
\end{figure}

Annotations, in the form of binary masks where 0 corresponds to the background and 1 represents a pixel belonging to the ROI, cover six organs: left and right lung, heart, trachea, esophagus, and spinal cord. 
Figure \ref{fig:pie} illustrates the ratio between background and ROIs, demonstrating that only a small portion of the scans is actually part of the segmentation target. Expert radiologists created the annotations. As CT scans are inherently volumetric, each scan consists of multiple slices. The dataset is split into 80\% for training, 10\% for validation, and 10\% for testing.

To isolate the effects of different training paradigms as much as possible and assess their impact on the final outcome, all models compared in this study are based on U-Net architecture~\cite{10.1007/978-3-319-24574-4_28}, including the generators of the GAN models.

    \subsection{Architecture of Adversarial Models} \label{sub:arch}
    The GAN models considered in this paper are based on~\cite{tan2021lgan} and share the same generator network, differing only in the design of the discriminator. The generator is a U-Net-based CNN that takes single slices of the CT scan as input (2D input) and outputs the segmentation mask.
    
        \begin{figure}
            \centering
            \includegraphics[width=\columnwidth]{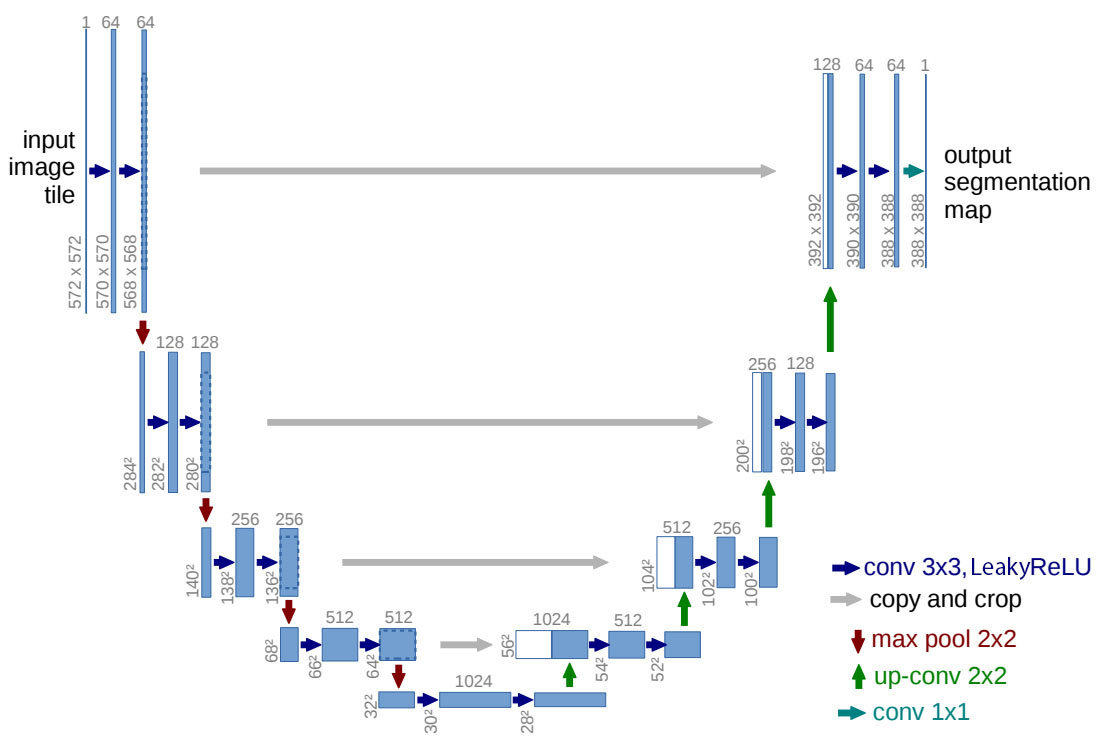}
            \caption{The architecture of the GANs generator}
            \label{fig:gen}
        \end{figure}
        
        In Figure~\ref{fig:gen}, a scheme showing the architecture of the generator is used. An important difference with the respect to the one used in the original paper is the use of Leaky ReLU activation functions instead of regular ReLU. Accordingly, also the loss function is different and reflects the discriminator used. It is, however, always comprised of two terms, one representing the generator's (G) and the other the discriminator's (D) loss: 

        Figure~\ref{fig:gen} depicts the architecture of the generator. 
        One significant deviation from the solution in~\cite{tan2021lgan} is the employment of Leaky ReLU activation functions 
        rather than regular ReLU. Furthermore, the loss function has been adapted to the discriminators used, despite it still comprises two terms, one representing the generator's (G) loss and the other representing the discriminator's (D) loss:
        
        \begin{equation}
		 	\mathcal{L}(x)=BCE[G(x), GT(x)] - \mathcal{L}_D(G(x), D(\cdot), x)
			\label{eq:loss}
		\end{equation}

        where $x$ is the input image, $GT(x)$ is the ground truth (the manual segmentation mask) of the input image $x$, $\mathcal{L}_D$ is the discriminator loss, and binary Cross Entropy (BCE) was chosen as generator loss.  

        We used three different discriminators to gain a more comprehensive understanding of the capabilities of adversarial training.

        \begin{figure}
            \centering
            \includegraphics[width=\columnwidth]{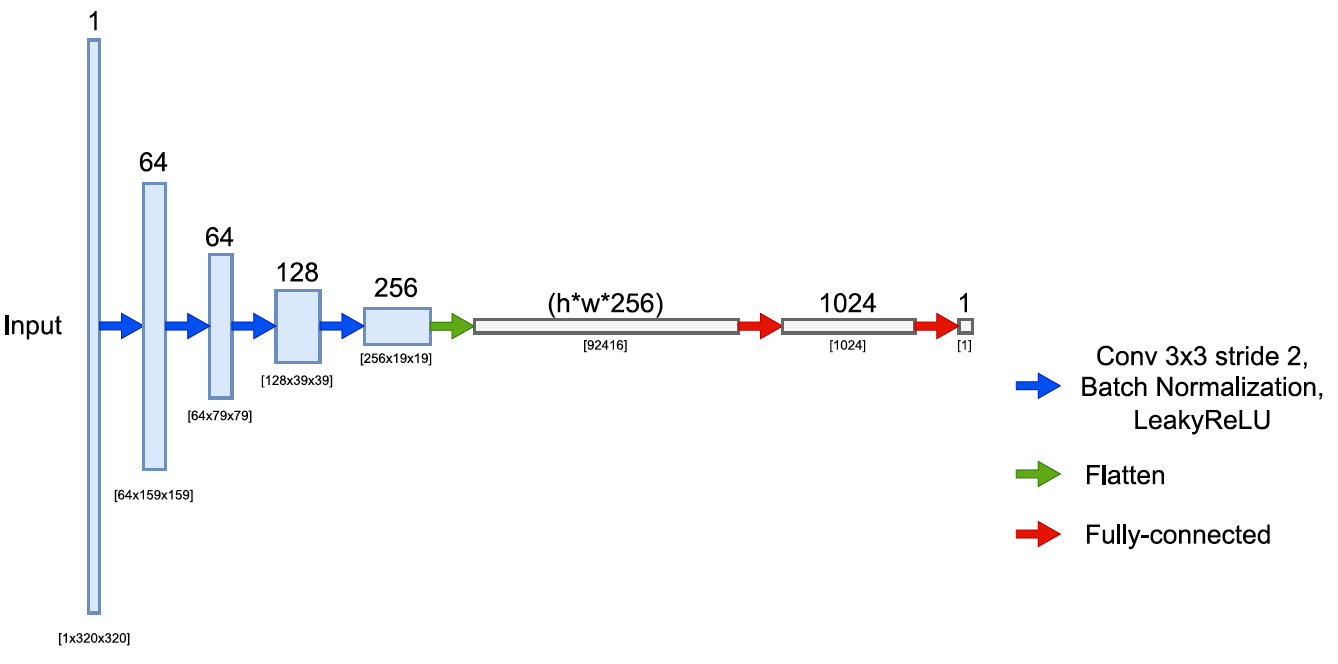}
            \caption{The architecture of the Product Discriminator}
            \label{fig:dis1}
        \end{figure}

        \begin{figure}
            \centering
            \includegraphics[width=\columnwidth]{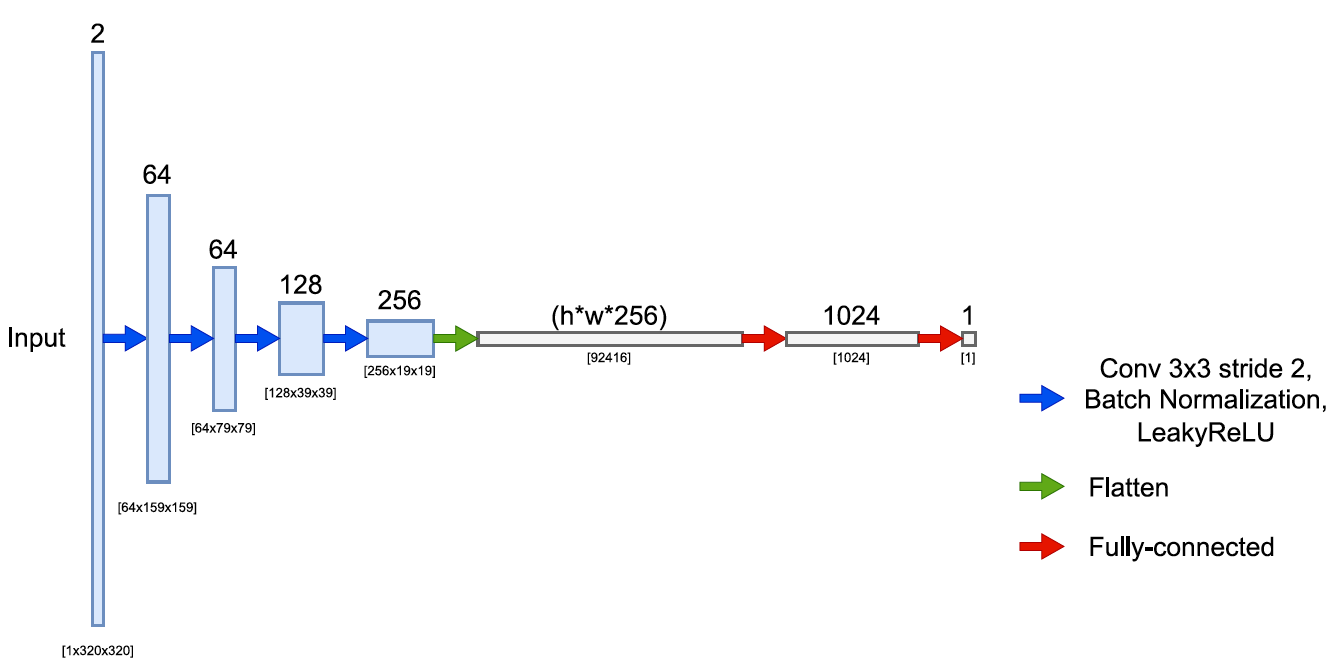}
            \caption{The architecture of the Early Fusion Discriminator}
            \label{fig:dis2}
        \end{figure}

        \begin{figure}
            \centering
            \includegraphics[width=\columnwidth]{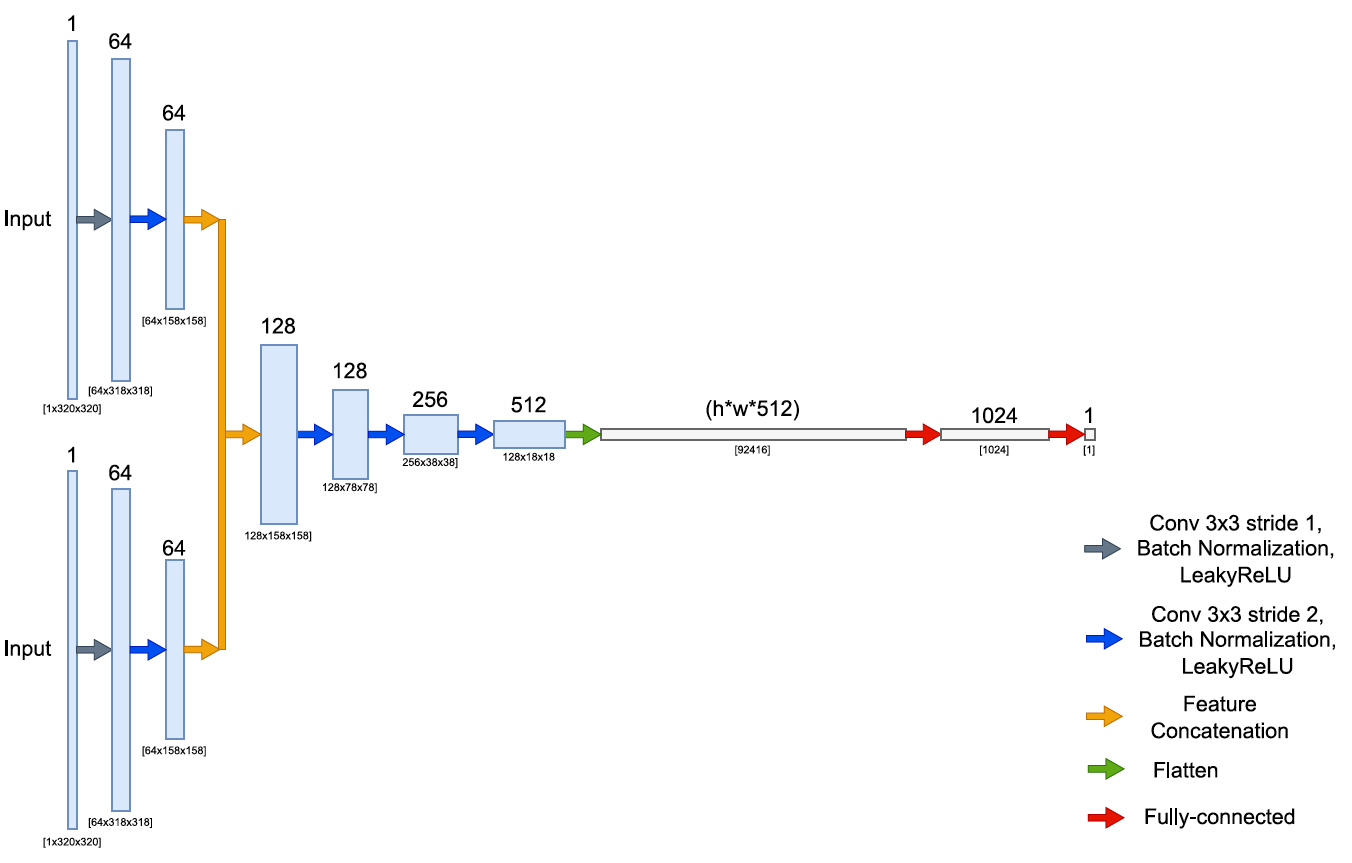}
            \caption{The architecture of the Late Fusion Discriminator}
            \label{fig:dis3}
        \end{figure}
        
        \medskip\noindent
        \textbf{Product Discriminator.}
        This discriminator takes as input the CT slice multiplied pixel-by-pixel by the corresponding segmentation mask. 
        Each pixel of the mask is encoded in the $[-1,1]$ range instead of the $[0,1]$ range as in~\cite{tan2021lgan}.
        To achieve this encoding, a $tanh$ activation function is applied to the output of the generator before multiplying it pixel-wise to the CT slice.
        The advantage of this encoding is twofold: it streamlines the training process and reduces the likelihood of converging to a trivial solution, as the information from the original image is not entirely overshadowed by the non-ROI pixels of the mask. 
        Figure \ref{fig:dis1} provides details of the discriminator's architecture. 
        The loss function for this discriminator is computed as:
        

            \begin{equation}
    		 	\mathcal{L}_D(G(x), D(.), x) = |E[D(G(x)\cdot x)]-E[D(GT(x)\cdot x)]|
    			\label{eq:dis1}
            \end{equation}

         \medskip\noindent
        \textbf{Early Fusion Discriminator.}
        This discriminator is based on the architecture shown in Figure \ref{fig:dis2}. It is similar to the previous one, with the key difference being that the input consists of two channels: one for the CT slice and the other for the corresponding segmentation mask. The pixel values of the mask are encoded in the $[0,1]$ range by applying a sigmoid activation function to the generator's output before using it as input. The pixel values of the CT slice are also in the range $[0,1]$, as the images have been normalized during preprocessing. The loss function for this discriminator is computed as:

            \begin{equation}
    		 	\mathcal{L}_D(G(x), D(\cdot), x) = |E[D(G(x)\circledast x)]-E[D(GT(x)\circledast x)]|
    			\label{eq:dis2}
            \end{equation}

        \medskip\noindent
        \textbf{Late Fusion Discriminator.}
        This discriminator takes as input both the CT slice and the segmentation mask, like the Early Fusion Discriminator, but processes them through the network in two parallel branches, which are combined after three convolutional blocks (hence, \textit{late} fusion). 
        Figure \ref{fig:dis1} provides details of the discriminator's architecture. 
        The loss function for this discriminator is computed as:
        
            \begin{equation}
    		 	\mathcal{L}_D(G(x), D(.), x) = |E[D(G(x), x)]-E[D(GT(x), x)]|
    			\label{eq:dis3}
            \end{equation}
            
        Two of the most popular segmentation models, which can currently be considered standard, have been selected for a fair comparison with the GANs:
        \begin{itemize}
            \item \textbf{SE-ResUNet}, introduced in \cite{cao2021cascaded}, utilizes a 2D U-Net backbone and incorporates Squeeze and Excitation blocks \cite{hu2018squeeze}. These blocks have been proven to provide a better feature representation. SE-ResUNet was specifically designed for thoracic organ segmentation and won the first StructSeg challenge. 
            \item \textbf{DeepLabV3} \cite{chen2017rethinking} is the latest iteration of the DeepLab architecture \cite{chen2017deeplab}. Its distinctive feature is the use of atrous convolutions, which enable the network to capture and utilize features at various scales within the image.
        \end{itemize}

    Both these networks have been trained from scratch as described in the following section.

    \subsection{Training and Experiments}
    Two experiments have been conducted:
        \begin{itemize}
        \item \textbf{Binary segmentation}, where a separate binary network is trained for each organ. 
        Each network takes a slice as input and returns a binary mask that distinguishes between the region of interest (ROI) and the background.
        
        \item \textbf{Multi-class ensemble segmentation}, in which multiple binary networks are used up until their last layer before the sigmoid or hyperbolic tangent activation function. The logit maps at this point can be interpreted as a probability map for pixels belonging or not belonging to the corresponding target class. Logit maps are stacked, and for each pixel, the maximum value is selected and assigned to the corresponding organ. In this manner, the final image has a value of 0 for the background and values from 1 to 6 for the pixels assigned to a specific organ. Two ensembles are created, one with the best-performing GANs and the other with the best model between SE-ResUNet and DeepLabV3. Models are chosen based on the highest validation scores obtained during training, ensuring that there is no bias on the test set.
        \end{itemize}
        
    To isolate the effects of the training paradigm as much as possible, the training process has been kept as similar as possible for both supervised and adversarial trained networks. The following training details apply to both the GANs and the supervised models: early stopping (with patience of 14 epochs) is used to halt training when a plateau in the validation loss is reached, along with a learning rate scheduler that decreases the learning rate by a factor of $0.2$ with patience of 10 epochs, starting from $10^{-5}$. 
    The Adam algorithm \cite{kingma2014adam} is employed to update the weights, with the following parameters: $\beta_1=0.5$, $\beta_2=0.999$, and weight decay$=5\cdot 10^{-4}$; the chosen batch size is 8.
    The models have been developed using the PyTorch library and trained on a machine equipped with an NVIDIA Quadro RTX 6000 GPU with CUDA 11.4.
\section{Results} \label{sec:res}
In this section, we present the results achieved by the binary segmentation models trained and by the multi-class ensemble segmentation. 
The performance evaluation relies on widely recognized metrics in the literature. 
The Dice Similarity Coefficient (DSC) assesses the accuracy of the overlap between the ground truth and the model segmentation. 
On the other hand, the 95th percentile of the Hausdorff Distance (HD95) quantifies the distance between the ground truth and the model segmentation. 
It is important to note that the DSC is calculated based on the entire volume for each patient, while the HD95 is computed per slice and subsequently averaged for each individual patient.

\subsection{Binary Segmentation} \label{sub:bin}

\begin{table*}
\centering
\resizebox{\textwidth}{!}{%
\begin{tabular}{|l|l|r|r|l|r|r|}
\hline
\rowcolor[HTML]{333333} 
\multicolumn{1}{|c|}{\cellcolor[HTML]{333333}{\color[HTML]{FFFFFF} \textbf{Target}}} & \multicolumn{1}{c|}{\cellcolor[HTML]{333333}{\color[HTML]{FFFFFF} \textbf{Model}}} & \multicolumn{1}{c|}{\cellcolor[HTML]{333333}{\color[HTML]{FFFFFF} \textbf{DSC}}} & \multicolumn{1}{c|}{\cellcolor[HTML]{333333}{\color[HTML]{FFFFFF} \textbf{HD95}}} & \multicolumn{1}{c|}{\cellcolor[HTML]{333333}{\color[HTML]{FFFFFF} \textbf{Target}}}                                              & \multicolumn{1}{c|}{\cellcolor[HTML]{333333}{\color[HTML]{FFFFFF} \textbf{DSC}}} & \multicolumn{1}{c|}{\cellcolor[HTML]{333333}{\color[HTML]{FFFFFF} \textbf{HD95}}} \\ \hline
\cellcolor[HTML]{680100}{\color[HTML]{FFFFFF} }                                      & GAN-Prod D                                                                         & \cellcolor[HTML]{FFFFC7}\textbf{0.9704}                                          & \cellcolor[HTML]{FFFFC7}\textbf{1.1293}                                           & \cellcolor[HTML]{963400}{\color[HTML]{FFFFFF} }                                                                                  & 0.9727                                                                           & 1.3071                                                                            \\
\cellcolor[HTML]{680100}{\color[HTML]{FFFFFF} }                                      &                                                                                    & \cellcolor[HTML]{FFFFC7}(0.9563, 0.9959)                                         & \cellcolor[HTML]{FFFFC7}(0.0000, 1.7321)                                          & \cellcolor[HTML]{963400}{\color[HTML]{FFFFFF} }                                                                                  & (0.9525, 0.9951)                                                                 & (1.0000, 2.0000)                                                                  \\ \cline{2-4} \cline{6-7} 
\cellcolor[HTML]{680100}{\color[HTML]{FFFFFF} }                                      & GAN-Early fus.                                                                     & 0.9698                                                                           & 1.2389                                                                            & \cellcolor[HTML]{963400}{\color[HTML]{FFFFFF} }                                                                                  & 0.9736                                                                           & \cellcolor[HTML]{FFFFC7}\textbf{1.2657}                                           \\
\cellcolor[HTML]{680100}{\color[HTML]{FFFFFF} }                                      &                                                                                    & (0.9552, 0.9956)                                                                 & (0.0000, 2.0000)                                                                  & \cellcolor[HTML]{963400}{\color[HTML]{FFFFFF} }                                                                                  & (0.9538, 0.9949)                                                                 & \cellcolor[HTML]{FFFFC7}(1.0000, 2.0000)                                          \\ \cline{2-4} \cline{6-7} 
\cellcolor[HTML]{680100}{\color[HTML]{FFFFFF} }                                      & GAN-Late fus.                                                                      & 0.9678                                                                           & 1.3803                                                                            & \cellcolor[HTML]{963400}{\color[HTML]{FFFFFF} }                                                                                  & \cellcolor[HTML]{FFFFC7}\textbf{0.9738}                                          & 1.3071                                                                            \\
\cellcolor[HTML]{680100}{\color[HTML]{FFFFFF} }                                      &                                                                                    & (0.9578, 0.9863)                                                                 & (1.0000, 2.0000)                                                                  & \cellcolor[HTML]{963400}{\color[HTML]{FFFFFF} }                                                                                  & \cellcolor[HTML]{FFFFC7}(0.9533, 0.9936)                                         & (1.0000, 2.0000)                                                                  \\ \cline{2-4} \cline{6-7} 
\cellcolor[HTML]{680100}{\color[HTML]{FFFFFF} }                                      & SE-ResUnet                                                                         & 0.9669                                                                           & 1.2828                                                                            & \cellcolor[HTML]{963400}{\color[HTML]{FFFFFF} }                                                                                  & 0.9711                                                                           & \cellcolor[HTML]{FFFFC7}\textbf{1.2657}                                           \\
\cellcolor[HTML]{680100}{\color[HTML]{FFFFFF} }                                      &                                                                                    & (0.9591, 0.9750)                                                                 & (1.0000, 2.0000)                                                                  & \cellcolor[HTML]{963400}{\color[HTML]{FFFFFF} }                                                                                  & (0.9600, 0.9779)                                                                 & \cellcolor[HTML]{FFFFC7}(1.0000, 2.0000)                                          \\ \cline{2-4} \cline{6-7} 
\cellcolor[HTML]{680100}{\color[HTML]{FFFFFF} }                                      & DeepLabV3                                                                          & 0.9652                                                                           & 1.3071                                                                            & \cellcolor[HTML]{963400}{\color[HTML]{FFFFFF} }                                                                                  & 0.9682                                                                           & 1.3632                                                                            \\
\multirow{-10}{*}{\cellcolor[HTML]{680100}{\color[HTML]{FFFFFF} \textbf{R-Lung}}}    &                                                                                    & (0.9565, 0.9727)                                                                 & (1.0000, 2.0000)                                                                  & \multirow{-10}{*}{\cellcolor[HTML]{963400}{\color[HTML]{FFFFFF} \textbf{L-Lung}}}                                                & (0.9572, 0.9748)                                                                 & (1.0000, 1.7321)                                                                  \\ \hline
\cellcolor[HTML]{CD9934}{\color[HTML]{FFFFFF} }                                      & GAN-Prod D                                                                         & 0.9231                                                                           & 2.8221                                                                            & \cellcolor[HTML]{036400}{\color[HTML]{FFFFFF} }                                                                                  & 0.8886                                                                           & \cellcolor[HTML]{FFFFC7}\textbf{1.5707}                                           \\
\cellcolor[HTML]{CD9934}{\color[HTML]{FFFFFF} }                                      &                                                                                    & (0.8440, 0.9636)                                                                 & (1.0000, 7.0711)                                                                  & \cellcolor[HTML]{036400}{\color[HTML]{FFFFFF} }                                                                                  & (0.8375, 0.9649)                                                                 & \cellcolor[HTML]{FFFFC7}(1.0000, 2.0000)                                          \\ \cline{2-4} \cline{6-7} 
\cellcolor[HTML]{CD9934}{\color[HTML]{FFFFFF} }                                      & GAN-Early fus.                                                                     & 0.9259                                                                           & 2.6585                                                                            & \cellcolor[HTML]{036400}{\color[HTML]{FFFFFF} }                                                                                  & 0.8861                                                                           & 1.6561                                                                            \\
\cellcolor[HTML]{CD9934}{\color[HTML]{FFFFFF} }                                      &                                                                                    & (0.8297, 0.9797)                                                                 & (1.0000, 6.6333)                                                                  & \cellcolor[HTML]{036400}{\color[HTML]{FFFFFF} }                                                                                  & (0.8375, 0.9663)                                                                 & (1.0000, 2.0000)                                                                  \\ \cline{2-4} \cline{6-7} 
\cellcolor[HTML]{CD9934}{\color[HTML]{FFFFFF} }                                      & GAN-Late fus.                                                                      & 0.9296                                                                           & 2.3887                                                                            & \cellcolor[HTML]{036400}{\color[HTML]{FFFFFF} }                                                                                  & 0.889                                                                            & 1.6025                                                                            \\
\cellcolor[HTML]{CD9934}{\color[HTML]{FFFFFF} }                                      &                                                                                    & (0.8392, 0.9721)                                                                 & (1.0000, 6.1644)                                                                  & \cellcolor[HTML]{036400}{\color[HTML]{FFFFFF} }                                                                                  & (0.8428, 0.9682)                                                                 & (1.0000, 2.0000)                                                                  \\ \cline{2-4} \cline{6-7} 
\cellcolor[HTML]{CD9934}{\color[HTML]{FFFFFF} }                                      & SE-ResUnet                                                                         & 0.8965                                                                           & 4.9144                                                                            & \cellcolor[HTML]{036400}{\color[HTML]{FFFFFF} }                                                                                  & \cellcolor[HTML]{FFFFC7}\textbf{0.892}                                           & 1.5901                                                                            \\
\cellcolor[HTML]{CD9934}{\color[HTML]{FFFFFF} }                                      &                                                                                    & (0.7441, 0.9528)                                                                 & (1.4142, 20.616)                                                                  & \cellcolor[HTML]{036400}{\color[HTML]{FFFFFF} }                                                                                  & \cellcolor[HTML]{FFFFC7}(0.8381, 0.9395)                                         & (1.0000, 2.0000)                                                                  \\ \cline{2-4} \cline{6-7} 
\cellcolor[HTML]{CD9934}{\color[HTML]{FFFFFF} }                                      & DeepLabV3                                                                          & \cellcolor[HTML]{FFFFC7}\textbf{0.9325}                                          & \cellcolor[HTML]{FFFFC7}\textbf{2.2061}                                           & \cellcolor[HTML]{036400}{\color[HTML]{FFFFFF} }                                                                                  & 0.8735                                                                           & 1.6439                                                                            \\
\multirow{-10}{*}{\cellcolor[HTML]{CD9934}{\color[HTML]{FFFFFF} \textbf{Heart}}}     &                                                                                    & \cellcolor[HTML]{FFFFC7}(0.8563, 0.9678)                                         & \cellcolor[HTML]{FFFFC7}(1.0000, 4.5826)                                          & \multirow{-10}{*}{\cellcolor[HTML]{036400}{\color[HTML]{FFFFFF} \textbf{\begin{tabular}[c]{@{}l@{}}Spinal\\ Cord\end{tabular}}}} & (0.8333, 0.9136)                                                                 & (1.0000, 2.0000)                                                                  \\ \hline
\cellcolor[HTML]{34696D}{\color[HTML]{FFFFFF} }                                      & GAN-Prod D                                                                         & 0.8319                                                                           & 3.0047                                                                            & \cellcolor[HTML]{340096}{\color[HTML]{FFFFFF} }                                                                                  & 0.7251                                                                           & 3.3277                                                                            \\
\cellcolor[HTML]{34696D}{\color[HTML]{FFFFFF} }                                      &                                                                                    & (0.7420, 0.9418)                                                                 & (1.0000, 6.8150)                                                                  & \cellcolor[HTML]{340096}{\color[HTML]{FFFFFF} }                                                                                  & (0.5022, 0.9505)                                                                 & (1.0000, 7.8740)                                                                  \\ \cline{2-4} \cline{6-7} 
\cellcolor[HTML]{34696D}{\color[HTML]{FFFFFF} }                                      & GAN-Early fus.                                                                     & 0.8419                                                                           & \cellcolor[HTML]{FFFFC7}\textbf{2.5856}                                           & \cellcolor[HTML]{340096}{\color[HTML]{FFFFFF} }                                                                                  & \cellcolor[HTML]{FFFFC7}\textbf{0.7587}                                          & \cellcolor[HTML]{FFFFC7}\textbf{2.9661}                                           \\
\cellcolor[HTML]{34696D}{\color[HTML]{FFFFFF} }                                      &                                                                                    & (0.7500, 0.9384)                                                                 & \cellcolor[HTML]{FFFFC7}(1.0000, 5.0990)                                          & \cellcolor[HTML]{340096}{\color[HTML]{FFFFFF} }                                                                                  & \cellcolor[HTML]{FFFFC7}(0.5823, 0.9610)                                         & \cellcolor[HTML]{FFFFC7}(1.0000, 7.7715)                                          \\ \cline{2-4} \cline{6-7} 
\cellcolor[HTML]{34696D}{\color[HTML]{FFFFFF} }                                      & GAN-Late fus.                                                                      & \cellcolor[HTML]{FFFFC7}\textbf{0.8482}                                          & 2.682                                                                             & \cellcolor[HTML]{340096}{\color[HTML]{FFFFFF} }                                                                                  & 0.7394                                                                           & 3.2503                                                                            \\
\cellcolor[HTML]{34696D}{\color[HTML]{FFFFFF} }                                      &                                                                                    & \cellcolor[HTML]{FFFFC7}(0.7265, 0.9651)                                         & (1.0000, 5.4772)                                                                  & \cellcolor[HTML]{340096}{\color[HTML]{FFFFFF} }                                                                                  & (0.5527, 0.9468)                                                                 & (1.0000, 8.5440)                                                                  \\ \cline{2-4} \cline{6-7} 
\cellcolor[HTML]{34696D}{\color[HTML]{FFFFFF} }                                      & SE-ResUnet                                                                         & 0.7822                                                                           & 5.0376                                                                            & \cellcolor[HTML]{340096}{\color[HTML]{FFFFFF} }                                                                                  & 0.6799                                                                           & 4.6414                                                                            \\
\cellcolor[HTML]{34696D}{\color[HTML]{FFFFFF} }                                      &                                                                                    & (0.6491, 0.8716)                                                                 & (2.0000, 9.0000)                                                                  & \cellcolor[HTML]{340096}{\color[HTML]{FFFFFF} }                                                                                  & (0.5883, 0.7665)                                                                 & (2.2361, 13.9284                                                                  \\ \cline{2-4} \cline{6-7} 
\cellcolor[HTML]{34696D}{\color[HTML]{FFFFFF} }                                      & DeepLabV3                                                                          & 0.7894                                                                           & 4.373                                                                             & \cellcolor[HTML]{340096}{\color[HTML]{FFFFFF} }                                                                                  & 0.7122                                                                           & 3.268                                                                             \\
\multirow{-10}{*}{\cellcolor[HTML]{34696D}{\color[HTML]{FFFFFF} \textbf{Trachea}}}   &                                                                                    & (0.6714, 0.8544)                                                                 & (1.4142, 15.0000)                                                                 & \multirow{-10}{*}{\cellcolor[HTML]{340096}{\color[HTML]{FFFFFF} \textbf{Esophagus}}}                                             & (0.5652, 0.8505)                                                                 & (1.4142, 5.3852)                                                                  \\ \cline{2-4} \cline{6-7} 
\end{tabular}}
\caption{Results for each binary model, OAR-wise; in brackets the maximum and minimum values are reported, to show the overall range; in bold and highlighted, the highest value for each OAR is highlighted.}
\label{tab:sin}
\end{table*}

    Table~\ref{tab:sin} displays the results for each OAR and binary model, highlighting the best performances in terms of DSC and HD95. 
    The results show that the model with the best performance according to DSC does not always result in the best HD95 score, suggesting that pixel-level accuracy is not necessarily related to the best delineation accuracy. 
    However, the model with the highest DSC score often achieves also very high HD95 score (see for the performances on the left lung, the spinal cord, and the trachea in Table~\ref{tab:sin}).
    Concerning the comparison between the different models, it is difficult to identify a clear overall winner, as each model excels in a specific organ (for instance, in Table~\ref{tab:sin} SE-ResUNet outperforms DeepLabV3 on the spinal cord, but is topped by DeepLabV3 on the heart and esophagus).
    In most of the organs, the GAN-based models perform similarly or better to the CNN-based models, i.e.,  Se-ResUnet and DeepLabV3.
    In particular, the performance difference, in favor of GAN-based models, is more noticeable in the trachea and the esophagus, which are the most challenging organs to segment. 
    This might suggest a more accurate behavior of adversarially trained models when the ROI-to-background ratio is low, as illustrated in figure \ref{fig:pie}.
    However, none of the three GAN-based models proposed, i.e., Product Discriminator, Early Fusion, 
      and Late Fusion outperforms consistently the others on all six organs.

    \subsection{Multi-class Ensemble Segmentation}

\begin{table}
\resizebox{\linewidth}{!}{%
\begin{tabular}{l|c|c|}
\cline{2-3}
                                                                                 & \cellcolor[HTML]{333333}{\color[HTML]{FFFFFF} GAN Ensemble} & \cellcolor[HTML]{333333}{\color[HTML]{FFFFFF} CNN Ensemble} \\ \hline
\multicolumn{1}{|l|}{\cellcolor[HTML]{680100}{\color[HTML]{FFFFFF} Right Lung}}  & \cellcolor[HTML]{FFFFFF}\textbf{0.9698}                     & 0.9669                                                      \\ \hline
\multicolumn{1}{|l|}{\cellcolor[HTML]{963400}{\color[HTML]{FFFFFF} Left Lung}}   & \cellcolor[HTML]{FFFFFF}\textbf{0.9736}                     & 0.971                                                       \\ \hline
\multicolumn{1}{|l|}{\cellcolor[HTML]{34696D}{\color[HTML]{FFFFFF} Trachea}}     & \cellcolor[HTML]{FFFFFF}\textbf{0.8488}                     & 0.7822                                                      \\ \hline
\multicolumn{1}{|l|}{\cellcolor[HTML]{340096}{\color[HTML]{FFFFFF} Esophagus}}   & \cellcolor[HTML]{FFFFFF}\textbf{0.759}                      & 0.7183                                                      \\ \hline
\multicolumn{1}{|l|}{\cellcolor[HTML]{986536}{\color[HTML]{FFFFFF} Heart}}       & 0.9258                                                      & \cellcolor[HTML]{FFFFFF}\textbf{0.9325}                     \\ \hline
\multicolumn{1}{|l|}{\cellcolor[HTML]{036400}{\color[HTML]{FFFFFF} Spinal Cord}} & \cellcolor[HTML]{FFFFFF}0.8861                              & \cellcolor[HTML]{FFFFFF}\textbf{0.8942}                     \\ \hline
\multicolumn{1}{|c|}{\textbf{Mean}}                                              & \multicolumn{1}{c|}{\textbf{0.8937}}                        & \multicolumn{1}{c|}{0.8775}                                 \\ \hline
\end{tabular}}
\caption{The performances in DSC of the ensembles. The best performances are reported in bold.}
\label{tab:mul}
\end{table}%

        \begin{figure}
            \begin{subfigure}{.5\columnwidth}
                \includegraphics[width=\linewidth]{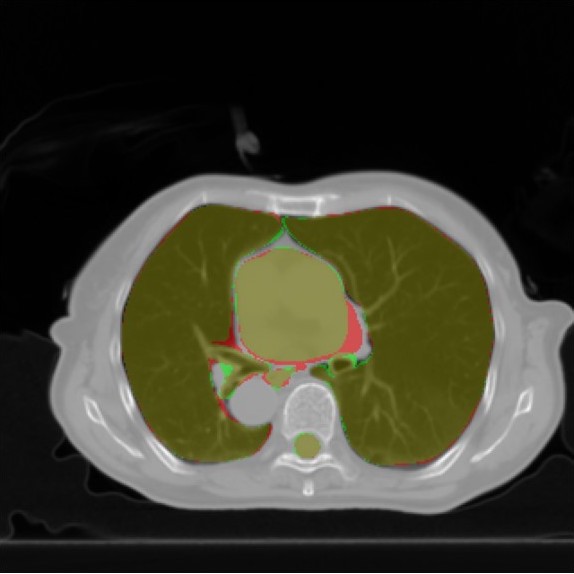}
                \caption{GAN ensemble}
                \label{fig:gan-ens}
            \end{subfigure}%
            \hspace*{\fill}
            \begin{subfigure}{.5\columnwidth}
                \includegraphics[width=\linewidth]{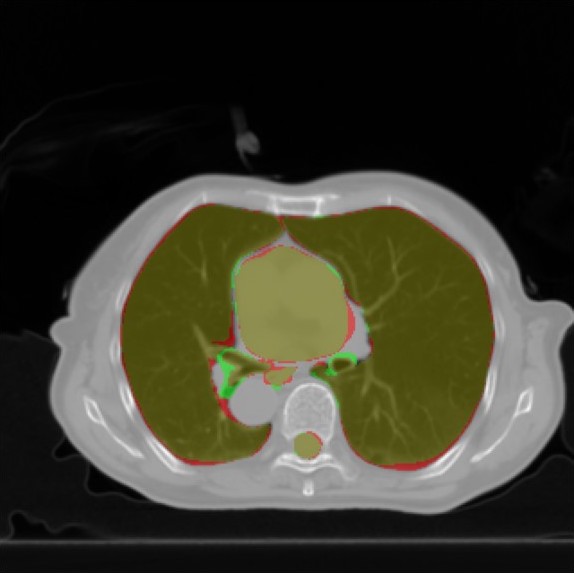}
                \caption{CNN ensemble}
                \label{fig:ref-ens}
            \end{subfigure}
            \caption{Visual comparison of the results of the ensembles on a sample slice. The \textcolor{olive}{yellow} areas identify the overlap between the GT and the output; \textcolor{green}{green} pixels highlight false positives; \textcolor{red}{red} pixels show false negatives.}
            \label{fig:ensemble}
        \end{figure}%

        Table \ref{tab:mul} shows the DSC scores achieved by the multi-class segmentation ensemble. 
        The results appear to be consistent with the findings previously discussed in Section~\ref{sub:bin}.
        GAN-based models perform similarly or slightly better than CNN-based ones, except for the two most challenging organs, i.e., trachea and esophagus, where the GAN-based approach seems to achieve more consistent performance.
        
        Figure~\ref{fig:ensemble} shows an example of the segmentation generated by the two models on the same slice, where it can be seen how the GAN-based model is less accurate around the heart (see the false negative area reported in red in Figure~\ref{fig:ensemble}) but more accurate in
        the rest of the slice.

        
        These findings offer valuable insights into the performance of adversarial and supervised learning paradigms. 
        While there is no definitive winner between the two approaches, GANs achieve comparable or marginally better performance across all organs. 
        This comparison can serve as a foundation for future research in the field of medical image segmentation, fostering further exploration of the advantages and limitations of these two learning paradigms. Ultimately, such research may contribute to the development of more accurate and robust segmentation models, enhancing their potential for use in clinical applications and improving patient care.
\section{Conclusions}
In this work, we compared GAN-based and CNN-based models for OAR segmentation in CT images. 
Specifically, we selected 2 state-of-the-art CNNs and proposed
  3 GANs models based on the architectures introduced in~\cite{tan2021lgan}.
We trained and compared these models on the StructSeg dataset, which consists of 50 thoracic CT series, annotated with contours of six organs (the two lungs, the heart, the trachea, the spinal cord, and the esophagus).
    
Our results show that GAN-based models are as effective as CNN-based ones in segmenting CT images, consistently achieving comparable or superior performance. 
Notably, GAN models exhibit a promising advantage when segmenting the most challenging targets, where they generally outperform their counterparts. 
This suggests that GAN models might successfully compete with state-of-the-art 
 CNN models on OAR segmentation.

Future research should explore this direction further, investigating whether different generators could lead to improved results. Additionally, the focus should shift toward not only performance but also usability, as a minimal improvement in DSC may not be relevant in real-world applications.

\section*{Acknowledgment}
This paper is supported by the FAIR (Future Artificial Intelligence Research) project, funded by the NextGenerationEU program within the PNRR-PE-AI scheme (M4C2, investment 1.3, line on Artificial Intelligence).

\bibliography{bibliography}
\bibliographystyle{IEEEtran}
\end{document}